\documentclass[fontsize=10pt]{article}
\usepackage{graphicx}
\usepackage{epstopdf, epsfig}
\usepackage{amsmath,esint}
\usepackage{amsmath,amsfonts,amsthm}
\usepackage{enumitem}
\usepackage{xcolor}
\usepackage{color}
\usepackage{soul}

\title{Exact second-order spatio-temporal structure-function relationships in non-stationary incompressible turbulent flows with Reynolds decomposition and phase averaging}

\author{Yisheng Zhang \thanks{Department of Civil and Mechancial Engineering, 2800 Kgs. Lyngby, Denmark; Currently at Alfa Laval Copenhagen A/S, 2860 Søborg, Denmark; yisheng.zhang.dk@gmail.com}
 \and Clara M. Velte  \thanks{Department of Civil and Mechancial Engineering, 2800 Kgs. Lyngby, Denmark}}

\begin{document}

\maketitle

\begin{abstract}
The Kármán–Howarth–Monin-Hill (KHMH) equation has been widely applied to scale-by-scale turbulent energy cascade studies in recent years; however, the forms and interpretations are not consistent. The present work generalizes to considering two different spatio-temporal points as a starting point to reformulate the KHMH equation based on Reynolds decomposition and phase averaging. The unaveraged form and phase averaged form are detailed and interpreted. Then the assumptions of homogeneity and isotropy are included in the KHMH equation to obtain the special form for homogeneous flows and isotropic flows.
\end{abstract}


\section{Introduction}

The mathematical foundation of the Richardson-Kolmogorov cascade based picture of turbulence is rooted in the evolution equation for the second-order structure function with an implicit time averaging (\cite{Richardson1922, Kolmogorov1941a, Kolmogorov1941b, Kolmogorov1941c, Batchelor1982}). This evolution equation, commonly known as the isotropic version of the Kármán-Howarth equation, was initially developed by \cite{Karman1938}. Subsequently, Monin and Yaglom extended this formulation to encompass anisotropic flows (see p. 403 in Ref.~\cite{Monin1971}), and Frisch synthesized these developments into the Kármán-Howarth-Monin equation (see p. 79 in Ref.~\cite{Frisch1995}).

Given the inherent fluctuations in interscale transfer rates, particularly in turbulent flows, a fully generalized version of the Kármán-Howarth-Monin equation, referred to as KHMH, has been developed by \cite{Duchon2000, Hill2001, Hill2002}. This extension aims to uncover the underlying mechanisms driving energy transfer in turbulence beyond the constraints imposed by averaging operations. By delving into the intricacies of energy transfer processes, the KHMH equation provides valuable insights into the dynamics of turbulent cascades and enhances our understanding of stationary turbulence phenomena.

The application of the KHMH equation manifests in two distinct directions: one employs Reynolds decomposition, such as \cite{Danaila2012, Valente2015,Alves2017, Alves2020, Zhou2020, Yao2022, Yao2023}, while the other operates without such decomposition, c.f. \cite{Danaila2001,Yasuda2018,Chen2022}. Despite the absence of an official definition of turbulence, statistical mechanics focuses on the residual effects post-averaging, which characterize turbulence. Consequently, the Reynolds decomposition form of the KHMH equation offers a comprehensive understanding of turbulent energy dynamics and the intricate interactions between mean flow and turbulent fluctuations. This approach provides a clearer and more intuitive depiction of turbulence dynamics, facilitating the analysis of energy transfer processes and the identification of coherent structures embedded within turbulent flows.

Inconsistencies regarding the formulations of the KHMH equation are evident across the literature. Appendix \ref{appA} delineates the identified variations in both the mathematical expressions and their corresponding interpretations. While the KHMH equation is stated without any explicit assumption in c.f. \cite{Yasuda2018}, it implicitly assumes a stationary flow regime, as the equation assumes that the two points involved share the same instance in time. However, in non-stationary flows, the equation should account for differences both in space and time between the two points due to acceleration and varying flow conditions. Therefore, a new formulation of the KHMH equation is necessary, one that incorporates both spatial and temporal variations between two points, to comprehensively analyze the scale-by-scale energy interactions in both space and time. This refined equation would provide a more accurate representation of turbulent dynamics in non-stationary flows, capturing the intricate spatiotemporal evolution of turbulent structures and energy transfer processes.

The overarching structure of this paper entails a systematic exploration of several key components. Initially, in Section 2, we embark on the derivation of the non-conventional two-point equations. Following this, Section 3 elucidates the manifestation of the scale-by-scale energy cascade equation in its phase-averaged manifestation. Subsequently, in Section 4, we invoke the fundamental assumptions of homogeneity and isotropy to facilitate our analysis. Section 5 then unveils an alternative representation of the scale-by-scale energy cascade equation. Finally, our deliberations culminate in Section 6, where a comprehensive discussions and conclusive remarks are presented.

\section{Exact unaveraged two-point equations with Reynolds decomposition} \label{sec:unaverage}

\subsection{Two-point Navier-Stokes equation with Reynolds decomposition}
The Navier-Stokes equation for velocity component $u_i (\boldsymbol{x})$ is
\begin{equation}\label{eq:NSE}
    \frac{\partial u_i (\boldsymbol{x})}{\partial t} + u_k (\boldsymbol{x}) \frac{\partial u_i (\boldsymbol{x})}{\partial x_k} = - \frac{\partial P (\boldsymbol{x})}{\partial x_i} + \nu \frac{\partial^2 u_i (\boldsymbol{x})}{(\partial x_k)^2}
\end{equation}
where $\boldsymbol{x}\in \mathbb{R}^4$ is a four-dimensional spatiotemporal vector, $P (\boldsymbol{x})$ is the pressure divided by the density and $\nu$ is the kinematic viscosity. Tensor notation is implied in Eq.~\eqref{eq:NSE}.

Considering two independent points $\boldsymbol{x}^+$ and $\boldsymbol{x}^- $ sharing the same dynamic system in Eq.~\eqref{eq:NSE}, the coordinate variables are $\boldsymbol{x}^+$ and $\boldsymbol{x}^-$. We denote $\tilde{u}^+_i=u_i (\boldsymbol{x}^+)$, $\tilde{u}^-_i=u_i (\boldsymbol{x}^-)$, $\tilde{P}^+_i=P_i (\boldsymbol{x}^+)$, $\tilde{P}^-_i=u_i (\boldsymbol{x}^-)$. The two-point equations are:

\begin{equation}\label{eq:twopoints}
    \begin{cases}
        \displaystyle\frac{\partial \tilde{u}_i^+}{\partial t^+} + \tilde{u}_k^+ \frac{\partial \tilde{u}_i^+}{\partial x^+_k} = - \frac{\partial \tilde{P}^+}{\partial x^+_i} + \nu \frac{\partial^2 \tilde{u}_i^+}{\partial x^+_k}\\[2ex] 
        \displaystyle \frac{\partial \tilde{u}_i^-}{\partial t^-} + \tilde{u}_k^- \frac{\partial \tilde{u}_i^-}{\partial x^-_k} = - \frac{\partial \tilde{P}^-}{\partial x^-_i} + \nu \frac{\partial^2 \tilde{u}_i^-}{\partial x^-_k}
    \end{cases}
\end{equation}

Reynolds decomposition is employed in the velocity and pressure with ensemble averaging that preserves the temporal dimension, $\tilde{u}^+=U^+_i + u^+_i$, $\tilde{u}^-_i=U^-_i + u^-_i$, $\tilde{P}^+=P^+ + p^+$ and $\tilde{P}^-=P^- + p^-$. Hence, the averaged terms, $U^+_i$, $U^-_i$, $P^+$ and $P^-$, include the variations in the temporal dimension. The two-point equation \eqref{eq:twopoints} is thus formulated as 

\begin{equation}\label{eq:twopointsRD}
    \begin{cases}
        \displaystyle\frac{\partial (U_i^+ + u_i^+)}{\partial t^+} + (U_k^+ + u_k^+)\frac{\partial (U_i^+ + u_i^+)}{\partial x^+_k} = - \frac{\partial (P^+ +p^+)}{\partial x^+_i} + \nu \frac{\partial^2 (U_i^+ + u_i^+)}{(\partial x^+_k)^2}\\[2ex] 
        \displaystyle \frac{\partial (U_i^- + u_i^-)}{\partial t^-} + (U_k^- + u_k^-) \frac{\partial (U_i^- + u_i^-)}{\partial x^-_k} = - \frac{\partial (P^-+p^-)}{\partial x^-_i} + \nu \frac{\partial^2 (U_i^- + u_i^-)}{(\partial x^-_k)^2}
    \end{cases}
\end{equation}

Leveraging the independent property detailed in Appendix~\ref{appB} Equation \eqref{eq:independent1}, along with the first line of the Reynolds decomposition \eqref{eq:twopointsRD}, we subtract $\tilde{u}^-$ and $\tilde{P}^-$ from their respective terms. Additionally, for the second line, we multiply by -1 and add $\tilde{u}^+$ and $\tilde{P}^+$ to the corresponding terms. This process yields the differences in velocity and pressure,

\begin{equation}\label{eq:RDdelta}
    \begin{cases}
        \displaystyle\frac{\partial (\delta U_i + \delta u_i)}{\partial t^+} + (U_k^+ + u_k^+)\frac{\partial (\delta U_i + \delta u_i)}{\partial x^+_k} = - \frac{\partial (\delta P +\delta p)}{\partial x^+_i} + \nu \frac{\partial^2 (\delta U_i + \delta u_i)}{(\partial x^+_k)^2}\\[2ex] 
        \displaystyle \frac{\partial (\delta U_i+ \delta u_i)}{\partial t^-} + (U_k^- + u_k^-) \frac{\partial (\delta U_i + \delta u_i)}{\partial x^-_k} = - \frac{\partial (\delta P+ \delta p)}{\partial x^-_i} + \nu \frac{\partial^2 (\delta U_i + \delta u_i)}{(\partial x^-_k)^2}
    \end{cases}
\end{equation}
where the difference of velocity and pressure denote $\delta U = U^+ - U^-$, $\delta u = u^+ - u^-$, $\delta P = P^+ - P^-$ and $\delta p = p^+ - p^-$.

\subsection{Variables with centre point and distance}
Following the methodology outlined in \cite{Hill2001, Hill2002}, we introduce new variables denoted as the center and distance of the two points, defined as $\boldsymbol{X} = (\boldsymbol{x}^+ + \boldsymbol{x}^-)/2$ and $\boldsymbol{r} = \boldsymbol{x}^+ - \boldsymbol{x}^-$. Here, $X_i$ and $r_i$ represent the components of the vectors $\boldsymbol{X}$ and $\boldsymbol{r}$ respectively, with $i\in\{0,1,2,3\}$, where $X_0$ and $r_0$ are temporal variables. The first-order derivative relations between the point position variables and the new variables are as follows:
\begin{equation}\label{eq:newvaribles}
    \begin{cases}
        \displaystyle \frac{\partial}{\partial x_i^+}=\frac{\partial}{\partial r_i} +\frac{1}{2}\frac{\partial}{\partial X_i}\\[2ex]
        \displaystyle \frac{\partial}{\partial x_i^-}=-\frac{\partial}{\partial r_i} +\frac{1}{2}\frac{\partial}{\partial X_i}
    \end{cases}
    ,\quad
    \begin{cases}
        \displaystyle \frac{\partial}{\partial X_i}=\frac{\partial}{\partial x^+_i} +\frac{\partial}{\partial x^-_i}\\[2ex]
        \displaystyle \frac{\partial}{\partial r_i}=\frac{1}{2}(\frac{\partial}{\partial x^+_i} - \frac{\partial}{\partial x^-_i})
    \end{cases}.
\end{equation}

The second-order derivative relations are
\begin{equation}\label{eq:2rdnewvaribles}
    \begin{cases}
        \displaystyle \frac{\partial^2}{(\partial x_i^+)^2}=\frac{\partial^2}{(\partial r_i)^2}+\frac{\partial^2}{\partial r_i\partial X_i} +\frac{1}{4}\frac{\partial^2}{(\partial X_i)^2}\\[2ex]
        \displaystyle \frac{\partial^2}{(\partial x_i^-)^2}=\frac{\partial^2}{(\partial r_i)^2}-\frac{\partial^2}{\partial r_i\partial X_i} +\frac{1}{4}\frac{\partial^2}{(\partial X_i)^2}
    \end{cases}.
\end{equation}

Equation \eqref{eq:newvaribles} and Equation \eqref{eq:2rdnewvaribles} are derived from Equation \eqref{appeq:M1} and Equation \eqref{appeq:M2} in Appendix \ref{appB}.

\subsection{Sum of the two equations with new variables}
The sum of Equations \eqref{eq:RDdelta} and the application of the property outlined in the continuous equation \eqref{appeq:nonlinear1} in Appendix \ref{appB} yields the following expression for the sum of the two-point equations:
\begin{eqnarray*}
        \displaystyle \left(\frac{\partial }{\partial t^+} + \frac{\partial }{\partial t^-}\right)(\delta U_i + \delta u_i)+ \frac{\partial (U_k^+ + u_k^+)(\delta U_i + \delta u_i) }{\partial x^+_k} +  \frac{\partial (U_k^- + u_k^-) (\delta U_i + \delta u_i)}{\partial x^-_k}\\[2ex]
        \displaystyle = - \left(\frac{\partial }{\partial x^+_i} +\frac{\partial }{\partial x^-_i} \right)(\delta P +\delta p) + \nu \left(\frac{\partial^2 }{(\partial x^+_k)^2}+\frac{\partial^2 }{(\partial x^-_k)^2}\right)(\delta U_i + \delta u_i)
\end{eqnarray*}

Substituting the variables from Equations \eqref{eq:newvaribles} into the equation above, we obtain:
\begin{eqnarray*}
        \displaystyle \frac{\partial }{\partial X_0} (\delta U_i + \delta u_i)+ \frac{\partial (U_k^+ + u_k^+)(\delta U_i + \delta u_i)}{\partial r_k} +\frac{1}{2}\frac{\partial (U_k^+ + u_k^+)(\delta U_i + \delta u_i)}{\partial X_k} \\[2ex]
       \displaystyle -\frac{\partial (U_k^- + u_k^-)(\delta U_i + \delta u_i)}{\partial r_k} +\frac{1}{2}\frac{\partial (U_k^- + u_k^-)(\delta U_i + \delta u_i)}{\partial X_k}  \\[2ex]
        = - \frac{\partial }{\partial X_i} (\delta P +\delta p) + \nu \left (\frac{\partial^2 }{(\partial x^+_k)^2}+\frac{\partial^2 }{(\partial x^-_k)^2}\right )(\delta U_i + \delta u_i).
\end{eqnarray*}

Combining the terms with the same variables, the result simplifies to:
\begin{eqnarray}
        \displaystyle \frac{\partial }{\partial X_0} (\delta U_i + \delta u_i)+ \frac{\partial (\delta U_k + \delta u_k)(\delta U_i + \delta u_i)}{\partial r_k} +\frac{\partial (U_k^* + u_k^*)(\delta U_i + \delta u_i)}{\partial X_k} \nonumber \\[2ex]
        = - \frac{\partial }{\partial X_i} (\delta P +\delta p) + \nu \left (\frac{\partial^2 }{(\partial x^+_k)^2}+\frac{\partial^2 }{(\partial x^-_k)^2} \right )(\delta U_i + \delta u_i),
        \label{eq:sum3}
\end{eqnarray}
where $U^*_i = (U^+_i + U^-_i)/2$ and $u^*_i = (u^+_i + u^-_i)/2$ are the averaged velocities.

\subsection{The two-point velocity distance correlation equation}
Multiplying equation \eqref{eq:sum3} by $\delta u_j$, the two-point velocity distance correlation equation is obtained,
\begin{eqnarray}
        \displaystyle \delta u_j\frac{\partial \delta U_i}{\partial X_0} + \delta u_j\frac{\partial \delta u_i}{\partial X_0} + \delta u_j \delta U_k \frac{\partial \delta U_i}{\partial r_k} + \delta u_j \delta u_k \frac{\partial \delta U_i }{\partial r_k} + \delta u_j \delta U_k \frac{\partial \delta u_i}{\partial r_k} \nonumber \\[2ex]
        + \delta u_j \delta u_k \frac{\partial \delta u_i}{\partial r_k} + \delta u_j U_k^* \frac{\partial \delta U_i }{\partial X_k}  +\delta u_j u_k^*\frac{\partial \delta U_i }{\partial X_k} + \delta u_j U_k^* \frac{\partial \delta u_i}{\partial X_k} + \delta u_j u_k^* \frac{\partial \delta u_i}{\partial X_k}\nonumber \\[2ex]
        = - \delta u_j\frac{\partial (\delta P + \delta p)}{\partial X_i}  + \nu \delta u_j \left (\frac{\partial^2 }{(\partial x^+_k)^2}+\frac{\partial^2 }{(\partial x^-_k)^2} \right )(\delta U_i + \delta u_i).
        \label{eq:summultiplyij}
\end{eqnarray}

When $j=i$ in equation \eqref{eq:summultiplyij}, both sides of the equation multiply by 2. Then the unaveraged scale-by-scale energy equation is obtained,
\begin{eqnarray}
        \displaystyle 2\delta u_i\frac{\partial \delta U_i}{\partial X_0} + \frac{\partial (\delta u_i)^2}{\partial X_0} + 2\delta u_i \delta U_k \frac{\partial \delta U_i}{\partial r_k} + 2\delta u_i \delta u_k \frac{\partial \delta U_i }{\partial r_k} + \delta U_k\frac{\partial (\delta u_i)^2}{\partial r_k} \nonumber \\[2ex]
        + \frac{\partial \delta u_k (\delta u_i)^2}{\partial r_k}  + 2\delta u_iU_k^* \frac{\partial \delta U_i }{\partial X_k}  + 2 \delta u_i u_k^*\frac{\partial \delta U_i }{\partial X_k} + 2 U_k^*\frac{\partial (\delta u_i)^2}{\partial X_k} + 2 \frac{\partial u_k^* (\delta u_i)^2}{\partial X_k}\nonumber \\[2ex]
        = -2 \frac{\partial (\delta u_i\delta P +\delta u_i\delta p)}{\partial X_i}  + 2\nu \delta u_i \left (\frac{\partial^2 }{(\partial x^+_k)^2}+\frac{\partial^2 }{(\partial x^-_k)^2}\right )(\delta U_i + \delta u_i).
        \label{eq:summultiplyii}
\end{eqnarray}

\subsection{The viscous term}
Utilizing Equation \eqref{appeq:viscosityij} Appendix \ref{appB} and the new variables \eqref{eq:newvaribles} and \eqref{eq:2rdnewvaribles}, the viscous term in Equation \eqref{eq:summultiplyij} can be expressed in terms of the new variables as follows:
\begin{eqnarray}
    &\displaystyle \nu \delta u_j \left[\frac{\partial^2 \delta u_i}{(\partial x^+_k)^2} + \frac{\partial^2 \delta u_i}{(\partial x^-_k)^2}\right] \nonumber \\[2ex]
    =& \displaystyle \nu \left[\frac{\partial^2 \delta u_i \delta u_j}{(\partial r_k)^2}+\frac{1}{4}\frac{\partial^2 \delta u_i \delta u_j}{(\partial X_k)^2}\right] -  \nu\frac{\partial \delta u_i}{\partial x^+_k}\frac{\partial \delta u_j}{\partial x^+_k} - \nu \frac{\partial \delta u_i}{\partial x^-_k}\frac{\partial \delta u_j}{\partial x^-_k}.
    \label{eq:viscocityij}
\end{eqnarray} 

Utilize \eqref{appeq:viscous} in the Appendix \ref{appB}, the viscous term in Equation \eqref{eq:summultiplyii} becomes
\begin{eqnarray}\label{eq:viscocityii}
    &\displaystyle 2\nu \delta u_i \left[\frac{\partial^2 \delta u_i}{(\partial x^+_k)^2} + \frac{\partial^2 \delta u_i}{(\partial x^-_k)^2}\right] \nonumber \\[2ex]
     =&\displaystyle 2 \nu \left[\frac{\partial^2 (\delta u_i)^2}{(\partial r_k)^2}+\frac{1}{4}\frac{\partial^2 (\delta u_i)^2}{(\partial X_k)^2}\right] -4\epsilon^* -2 \nu \Delta ( P^* + p^*) - 2\nu F^+_{ij} - 2\nu F^-_{ij} - 2 \nu H^+ \quad \quad 
\end{eqnarray}
where $\epsilon^*=(\epsilon^+ + \epsilon^-)/2$ is the average of the two-point dissipation rate, $\epsilon^+$ and $\epsilon^-$, and $F^+_{ij}$ and $F^-_{ij}$ are the two-point terms of $F_{ij}$ in Equation \eqref{appeq:viscous} in Appendix \ref{appB}, $H^+=\displaystyle\frac{\partial U^+_i}{\partial x^+_j}\frac{\partial U^+_j}{\partial x^+_i} + \frac{\partial U^-_i}{\partial x^-_j}\frac{\partial U^-_j}{\partial x^-_i}$. 

\subsection{The scale-by-scale equation}
Equation \eqref{eq:summultiplyii} replicated with the viscosity term \eqref{eq:viscocityii}:
\begin{eqnarray}
        \displaystyle \underbrace{2\delta u_i\frac{\partial \delta U_i}{\partial X_0}}_{\substack{\text{Stochastic Tran-} \\ \text{sient transport}}} + \underbrace{\frac{\partial (\delta u_i)^2}{\partial X_0}}_{\substack{\text{Transient } \\ \text{TKE transport}}} + \underbrace{ 2\delta u_i \delta U_k \frac{\partial \delta U_i}{\partial r_k}}_{\substack{\text{Stochastic } \\ \text{production}}} + \underbrace{2\delta u_i \delta u_k \frac{\partial \delta U_i }{\partial r_k}}_{\substack{\text{Turbulent produc-} \\ \text{tion by transfer}}} + \underbrace{\delta U_k\frac{\partial (\delta u_i)^2}{\partial r_k}}_{\substack{\text{Linear TKE} \\ \text{transfer}}} \nonumber \\[2ex]
        + \underbrace{ \frac{\partial \delta u_k(\delta u_i)^2}{\partial r_k}}_{\substack{\text{Nonlinear TKE} \\ \text{transfer}}} +\underbrace{2\delta u_iU_k^*\frac{\partial \delta U_i }{\partial X_k}}_{\substack{\text{Stochastic } \\ \text{transfer}}} + \underbrace{2\delta u_i u_k^*\frac{\partial \delta U_i }{\partial X_k}}_{\substack{\text{Turbulent production} \\ \text{by transport}}} +\underbrace{ U_k^*\frac{\partial (\delta u_i)^2}{\partial X_k}}_{\substack{\text{Linear TKE} \\ \text{transport}}} + \underbrace{ \frac{\partial u_k^*(\delta u_i)^2}{\partial X_k}}_{\substack{\text{Non linear TKE} \\ \text{transport}}} \nonumber \\[2ex]
        \displaystyle = - \underbrace{2\frac{\partial \delta u_i\delta P}{\partial X_i} }_{\substack{\text{Stochastic pres-} \\ \text{sure transport}}} -\underbrace{2\frac{\partial \delta u_i \delta p}{\partial X_i} }_{\substack{\text{Turbulent pres-} \\ \text{sure transport}}}+ \underbrace{2\nu \delta u_i \left[\frac{\partial^2 \delta U_i}{(\partial r_k)^2}+\frac{1}{4}\frac{\partial^2 \delta U_i}{(\partial X_k)^2}\right]}_{\substack{\text{Stochastic viscous} \\ \text{transfer}}} + \underbrace{2 \nu \frac{\partial^2 (\delta u_i)^2}{(\partial r_k)^2}}_{\substack{\text{Viscous diffusion} \\ \text{transfer}}}  \nonumber \\[2ex]
        \displaystyle + \underbrace{\frac{\nu}{2}\frac{\partial^2 (\delta u_i)^2}{(\partial X_k)^2}}_{\substack{\text{Viscous diffusion} \\ \text{transport}}} -\underbrace{4 \epsilon^*}_{\text{Dissipation} } - \underbrace{2\nu (\Delta P^* + H^+)}_{\substack{\text{Viscous diffusion} \\ \text{transport from mean}}} - \underbrace{2\nu F^+_{ij} -2\nu F^-_{ij} - 2\nu \Delta p^*}_{\substack{\text{Stochastic viscous} \\ \text{diffusion}}}
        \label{eq:summultiplyinter}
\end{eqnarray}
The term 'transfer' refers to the scale-by-scale transfer concerning the radius $\boldsymbol{r}$. The term 'transport' refers to the energy/structure function traveling to another position concerning the distance $\boldsymbol{X}$. It is noted that all stochastic terms equate to zero following the averaging process.

\section{Phase averaged equations}
In accordance with the findings of \cite{Zhang2023}, which highlight that the phase average is equivalent to the ensemble average in periodic flows, the phase average is applied to the scale-by-scale equation \eqref{eq:summultiplyinter} in this section. Moreover, due to the commutative property of the indices $i$ and $j$ within the summation, as exemplified by $\displaystyle\delta u_j\frac{\partial \delta p}{\partial X_i} = \delta u_i\frac{\partial \delta p}{\partial X_j}$, Equation \eqref{eq:summultiplyij} and \eqref{eq:viscocityij} are combined and phase averaged, then rearranged by commuting terms and then summed back to attain the following form:
\begin{eqnarray}
        \displaystyle  \frac{\partial \langle\delta u_i\delta u_j \rangle }{\partial X_0} + 2\langle \delta u_j \delta u_k \rangle \frac{\partial \delta U_i }{\partial r_k} +  \frac{\partial \langle\delta u_i\delta u_j \rangle \delta U_k}{\partial r_k} + \frac{\partial \langle \delta u_i\delta u_j \delta u_k \rangle}{\partial r_k}  \nonumber \\[2ex]
         +2\langle \delta u_j u_k^*\rangle\frac{\partial \delta U_i }{\partial X_k} + \frac{\partial \langle \delta u_i\delta u_j \rangle U_k^*}{\partial X_k} +  \frac{\partial \langle \delta u_i\delta u_j u_k^*\rangle}{\partial X_k} = - 2\langle\delta u_j\frac{\partial \delta p}{\partial X_i} \rangle \nonumber \\[2ex]
        + \displaystyle 2\nu \left[\frac{\partial^2 \langle \delta u_i \delta u_j \rangle}{(\partial r_k)^2}+\frac{1}{4}\frac{\partial^2 \langle \delta u_i \delta u_j\rangle}{(\partial X_k)^2}\right] -  2\nu\langle \frac{\partial \delta u_i}{\partial x^+_k}\frac{\partial \delta u_j}{\partial x^+_k} \rangle - 2\nu \langle \frac{\partial \delta u_i}{\partial x^-_k}\frac{\partial \delta u_j}{\partial x^-_k}\rangle.
        \label{eq:averagedij}
\end{eqnarray}

Using a similar process, Equation \eqref{eq:summultiplyinter} can be reformulated as:
\begin{eqnarray}
        \displaystyle \underbrace{\frac{\partial \langle (\delta u_i)^2 \rangle }{\partial X_0}}_{\substack{\mathcal{A}\text{ Transient } \\ \text{TKE transport}}} + \underbrace{\delta U_k\frac{\partial \langle (\delta u_i)^2 \rangle}{\partial r_k}}_{\substack{\mathcal{L} _1\text{ Linear TKE} \\ \text{transfer}}} + \underbrace{ U_k^*\frac{\partial \langle (\delta u_i)^2 \rangle}{\partial X_k}}_{\substack{\mathcal{L} _2 \text{ Linear TKE} \\ \text{transport}}} + \underbrace{\frac{\partial \langle (\delta u_i)^2 \delta u_k \rangle}{\partial r_k} }_{\substack{\mathcal{\Pi}_1 \text{Nonlinear TKE} \\ \text{transfer}}}  \nonumber \\[2ex]
        + \underbrace{ \frac{\partial \langle (\delta u_i)^2 u_k^* \rangle}{\partial X_k} }_{\substack{\mathcal{\Pi} _2 \text{Nonlinear TKE} \\ \text{transport}}} + \underbrace{2\langle \delta u_i \delta u_k  \rangle\frac{\partial \delta U_i }{\partial r_k}}_{\substack{\mathcal{F}_1 \text{ Turbulent produc-} \\ \text{tion by transfer}}} + \underbrace{\langle \delta u_i u_k^* \rangle\frac{\partial \delta U_i }{\partial X_k}}_{\substack{\mathcal{F}_2 \text{ Turbulent production} \\ \text{by transport}}}  \nonumber \\[2ex]
        \displaystyle = -\underbrace{2\frac{\partial \langle \delta u_i\delta p \rangle}{\partial X_i} }_{ \substack{\mathcal{P}\text{ Turbulent pers-} \\ \text{sure transport}}} + \underbrace{2 \nu \frac{\partial^2 \langle (\delta u_i)^2 \rangle}{(\partial r_k)^2}}_{\substack{\mathcal{D}_1\text{ Viscous diffusion} \\ \text{transfer}}} + \underbrace{\frac{\nu}{2}\frac{\partial^2 \langle (\delta u_i)^2 \rangle}{(\partial X_k)^2}}_{\substack{\mathcal{D}_2 \text{ Viscous diffusion} \\ \text{transport}}} -\underbrace{4 \langle \epsilon^* \rangle}_{\mathcal{D}_\epsilon \text{ Dissipation} } \nonumber \\[2ex]
        - \underbrace{2\nu (\Delta P^* - \frac{1}{2}\frac{\partial \delta U_i}{\partial X_j} \frac{\partial \delta U_j}{\partial X_i} - 2\frac{\partial \delta U_i}{\partial r_j} \frac{\partial \delta U_j}{\partial r_i})}_{\substack{\mathcal{D}_3 \text{Viscous diffusion} \\ \text{transport from mean flow}}}
        \label{eq:averagedsbs}
\end{eqnarray}

Equation \eqref{eq:averagedsbs} is the phase averaged scale-by-scale energy cascade equation without any assumption.
\begin{enumerate}
    \item The term $\mathcal{A} = \partial \langle (\delta u_i)^2 \rangle/\partial t$ represents the rate of change in time of averaged energy $\langle (\delta u_i)^2 \rangle$ at a given physical point $X_i$ and separation $r_i$. The the flow is stationary or temporal homogeneous, statistics of the averaged energy are invariant in time and therefore $\mathcal{A}$ vanishes.
    \item The term $\mathcal{L}_1 = \delta U_k(\partial \langle (\delta u_i)^2 \rangle)/(\partial r_k)$ is the linear energy interscale transfer term. This term represents the energy transfer of $\langle (\delta u_i)^2 \rangle$ in the scales $r_i$ by the mean flow. In fact, integration of  $\mathcal{L}_1$ over a spherical volume in $\boldsymbol{r}$ in physical space and use of Gauss’ theorem yields energetic flux with TKE and mean flow differences through the spherical volume. The energetic flux shows the turbulence cascade with the mean flow differences.
    \item The term $\mathcal{L}_2 = U^*_k(\partial \langle (\delta u_i)^2 \rangle)/(\partial X_k)$ is the linear energy transfer with averaged mean flow term. This term represents the energy $\langle (\delta u_i)^2 \rangle$ transport in the direction of the averaged mean flow $U^*_k$. If the flow is homogeneous, the term $\mathcal{L}_2$ vanishes.
    \item The term $\mathcal{\Pi}_1 =[\partial \langle (\delta u_i)^2 \delta u_k \rangle]/(\partial r_k)$ is the nonlinear interscale transfer rate and accounts for the effect of nonlinear interactions in redistributing $\langle (\delta u_i)^2 \rangle$ within the $\boldsymbol{r}$-space, and is given by the divergence in scale space of the flux $\langle (\delta u_i)^2 \delta u_k \rangle$. The surface integral of $\mathcal{\Pi}_1$ over a volume sphere $\boldsymbol{r}$ shows the triadic energy interaction with the Gauss's theorem, $\iiint\limits_{V(\boldsymbol{r})} (\nabla \cdot \langle (\delta u_i)^2 \delta u_k \rangle) d V = \oiint\limits_{S(\boldsymbol{r})} [\partial \langle (\delta u_i)^2 \delta u_k \rangle]/(\partial r_k) \cdot \hat{r_k}/\vert r \vert d S$, then the flux integral shows the turbulence cascade within the turbulent scales, corresponding to a length scale equal to the radius of the sphere.
    \item The term $\mathcal{\Pi}_2=[\partial \langle (\delta u_i)^2 u_k^* \rangle]/\partial X_k$ is the energy transport in physical space due to turbulent fluctuations. If the flow is homogeneous, the term $\mathcal{\Pi}_2$ vanishes.
    \item The term $\mathcal{F}_1=2(\langle \delta u_i \delta u_k  \rangle)\partial \delta U_i/ \partial r_k$ is the production of the averaged two-point correlation $\langle \delta u_i \delta u_k  \rangle$ by mean flow differences gradients in the scales. If the flow is homogeneous, the term $\langle \delta u_i \delta u_k  \rangle=0$ when $i\neq k$, which means $\mathcal{F}_1=2(\langle \delta u_k \delta u_k  \rangle)\partial \delta U_k/ \partial r_k$. If the flow is isotropic, $\langle \delta u_1 \delta u_1  \rangle=\langle \delta u_2 \delta u_2  \rangle=\langle \delta u_3 \delta u_3  \rangle$, the term $\mathcal{F}_1$ vanishes due to continuity.
    \item The term $\mathcal{F}_2=2(\langle \delta u_i u^*_k  \rangle)\partial \delta U_i/ \partial r_k$ is the production of the averaged two-point correlation $\langle \delta u_i u^*_k  \rangle$ by mean flow difference gradients in the physical space. If the flow is homogeneous, $\langle \delta u_i u^*_k  \rangle=\langle u^+_i u^+_k  \rangle-\langle u^-_i u^-_k  \rangle=0$, and $\mathcal{F}_2$ vanishes.
    \item The term $\mathcal{P}=2(\partial \langle \delta u_i\delta p \rangle)/(\partial X_i) $ is the correlation between velocity differences and pressure differences transport in physical space.
    \item The term $\mathcal{D}_1=2 \nu \partial^2 \langle (\delta u_i)^2 \rangle/(\partial r_k)^2 $ is the diffusion transfer in scale space by viscosity.
    \item The term $\mathcal{D}_2= \frac{1}{2}\nu \partial^2 \langle (\delta u_i)^2 \rangle/(\partial X_k)^2$ is the diffusion transport in physical space due to viscosity. This term is analogous to the diffusion term appearing in the single-point turbulent kinetic energy equation, and thus, its contribution is expected to be small. If the flow is homogeneous, the term $\mathcal{D}_2$ vanishes.
    \item The dissipation term $\mathcal{D}_\epsilon = 4 \langle \epsilon^* \rangle$ is related to the two-point average dissipation rate.
    \item The term $\mathcal{D}_3=2\nu [\Delta P^* -  \frac{1}{2}(\partial \delta U_i/\partial X_j)(\partial \delta U_j/\partial X_i) - 2(\partial \delta U_i/\partial r_j)(\partial \delta U_j/\partial r_i)]$ is transport from mean flow. If the flow is homogeneous, the term $\mathcal{D}_3$ vanishes (see Section 9.7.2 in \cite{George2013}).
\end{enumerate}
\section{Assumptions and the equation forms}

\subsection{The assumption of Homogeneity}
Under the assumption of homogeneity in turbulent flow and the invariance of spatial transport statistics for flow velocity and pressure, expressed as $\langle u^+_i \rangle = \langle u^-_i \rangle$, $\langle (u^+_i)^n (u^+_j)^m\rangle = \langle (u^-_i)^n (u^-_j)^m \rangle$, and $\partial \langle (\delta u_i)^n(\delta u_j)^m (\delta u_s)^l (\delta p)^h\rangle/\partial X_k = 0$, where $i,j,k,n,m,l,h\in\{ 0,1,2,3\}$, the terms $\mathcal{A},\mathcal{L}_2,\mathcal{\Pi}_1,\mathcal{F}_2, \mathcal{P}, \mathcal{D}_2$ vanish from Equation \eqref{eq:averagedsbs}. Consequently, the scale-by-scale equation can be formulated as:
\begin{eqnarray}
        \displaystyle  \delta U_k\frac{\partial \langle (\delta u_i)^2 \rangle}{\partial r_k}  + \frac{\partial \langle (\delta u_i)^2 \delta u_k \rangle}{\partial r_k} + 2\langle \delta u_i \delta u_k  \rangle\frac{\partial \delta U_i }{\partial r_k} = 2 \nu \frac{\partial^2 \langle (\delta u_i)^2 \rangle}{(\partial r_k)^2} -4 \langle \epsilon^* \rangle
        \label{eq:homogeneoussbs}
\end{eqnarray}

Expressing in the symbols of the terms:
\begin{eqnarray*}
    \mathcal{L}_1 + \mathcal{\Pi} _1 + \mathcal{F }_1 = \mathcal{D}_1 - \mathcal{D}_\epsilon 
\end{eqnarray*}
Note that homogeneity is assumed both in space and time. The terms $\mathcal{L}_2,\mathcal{\Pi}_1,\mathcal{F}_2, \mathcal{P}, \mathcal{D}_2, \mathcal{D}_3$ vanish due to spatial homogeneity, and the term $\mathcal{A}$ equates to zero due to temporal homogeneity (stationary).

The two-point correlation equation \eqref{eq:averagedij} reduces to 
\begin{eqnarray}
        \displaystyle  2\langle \delta u_j \delta u_k \rangle \frac{\partial \delta U_i }{\partial r_k} +  \frac{\partial \langle\delta u_i\delta u_j \rangle \delta U_k}{\partial r_k}         +  \frac{\partial \langle \delta u_i\delta u_j \delta u_k \rangle}{\partial r_k} +2\langle \delta u_j u_k^*\rangle\frac{\partial \delta U_i }{\partial X_k} \nonumber \\[2ex]
        = \displaystyle 2\nu \frac{\partial^2 \langle \delta u_i \delta u_j\rangle}{(\partial r_k)^2} -  \frac{4}{3}\langle \epsilon^* \rangle \delta_{ij}.
        \label{eq:homogeneousij}
\end{eqnarray}

It is important to note that the homogeneity in this paper is based on the statistical law of the fluctuation velocity components. If the homogeneity meant the non-Reynolds decomposition velocity (the total velocity), the mean velocity is homogeneous and $\delta U_i = 0$. Then scale-by-scale equation \eqref{eq:homogeneoussbs} is
\begin{eqnarray}
        \frac{\partial \langle (\delta u_i)^2 \delta u_k \rangle}{\partial r_k}  = 2 \nu \frac{\partial^2 \langle (\delta u_i)^2 \rangle}{(\partial r_k)^2} -4 \langle \epsilon^* \rangle,
        \label{eq:homogeneousisotropicsbs}
\end{eqnarray}
which is $\mathcal{\Pi} _1 = \mathcal{D}_1 - \mathcal{D}_\epsilon$. And the two-point correlation equation \eqref{eq:homogeneousij} is
\begin{eqnarray}
        \frac{\partial \langle \delta u_i\delta u_j \delta u_k \rangle}{\partial r_k}  = 2\nu \frac{\partial^2 \langle \delta u_i \delta u_j\rangle}{(\partial r_k)^2} -  \frac{4}{3}\langle \epsilon^* \rangle \delta_{ij},
        \label{eq:homogeneousisotropicsij}
\end{eqnarray}
which is the same with Monin and Yaglom's anisotropic form with $\delta u$ (see Equation 22.15 in \cite{Monin1971}). \cite{Monin1971} use the structure funcitions $D_{ijk}=\langle \delta u_i\delta u_j \delta u_k \rangle$ and $D_{ij}=\langle \delta u_i\delta u_j \rangle$ to derive the anisotropic form of the velocity structure equations, which is equivalent to Equation \eqref{eq:homogeneousisotropicsij}. Furthermore, under the assumption of local isotropy (see Sections 22.1 and 13.3 in \cite{Monin1971}), the velocity structure equation $\displaystyle D_{LLL}-6\nu\frac{dD_{LL}}{dr}=-\frac{4}{5}\langle \epsilon \rangle$ can be derived. Consequently, Kolmogorov's structure equation from \cite{Kolmogorov1941c} is also recovered. can be obtained by the local isotropy assumption (see section 22.1 and section 13.3 in \cite{Monin1971}). So the Kolmogorov's structure equation in \cite{Kolmogorov1941c} is also obtained.

However, this paper employs Reynolds decomposition and phase averaging over the temporal dimension. Here $\displaystyle \frac{\partial }{\partial r_0}$ and $\displaystyle \frac{\partial }{\partial X_0}$ represent time derivatives, which are assumed to be zero in steady flows.
\subsection{The assumption of isotropy}
The isotropy assumption is implied by the homogeneity, in \cite{Karman1938, Kolmogorov1941c, George2013}. Employed to Equation \eqref{eq:homogeneoussbs}, the scale-by-scale energy cascade equation reduces to 
\begin{equation}
        \delta U_k\frac{\partial \langle (\delta u_i)^2 \rangle}{\partial r_k} + \frac{\partial \langle (\delta u_i)^2 \delta u_k \rangle}{\partial r_k} = 2 \nu \frac{\partial^2 \langle (\delta u_i)^2 \rangle}{(\partial r_k)^2} - 4 \langle \epsilon^* \rangle
        \label{eq:isotropicsbs}
\end{equation}

Expressed in the terms of the symbols of the terms:
\begin{eqnarray*}
    \mathcal{L}_1 + \mathcal{\Pi} _1 = \mathcal{D}_1 - \mathcal{D}_\epsilon 
\end{eqnarray*}
Note that the production from mean flow term $\mathcal{F}_2$ vanishes due to the isotropic assumption.

The two-point correlation equation \eqref{eq:homogeneousij} is reduced to 
\begin{eqnarray}
        \displaystyle  2\langle \delta u_j \delta u_k \rangle \frac{\partial \delta U_i }{\partial r_k} +  \frac{\partial \langle\delta u_i\delta u_j \rangle \delta U_k}{\partial r_k}         +  \frac{\partial \langle \delta u_i\delta u_j \delta u_k \rangle}{\partial r_k} \nonumber \\[2ex]
        = \displaystyle 2\nu \frac{\partial^2 \langle \delta u_i\delta u_j \rangle}{(\partial r_k)^2} -  2\nu\langle \frac{\partial \delta u_i}{\partial x^+_k}\frac{\partial \delta u_j}{\partial x^+_k} \rangle - 2\nu \langle \frac{\partial \delta u_i}{\partial x^-_k}\frac{\partial \delta u_j}{\partial x^-_k}\rangle.
        \label{eq:isotropicij}
\end{eqnarray}

When $i\neq j$, $\langle \delta u_i \delta u_j \rangle =0 $, so Equation \eqref{eq:isotropicij} becomes
\begin{eqnarray}
        \displaystyle  2\langle \delta u_j \delta u_k \rangle \frac{\partial \delta U_i }{\partial r_k}  +  \frac{\partial \langle \delta u_i\delta u_j \delta u_k \rangle}{\partial r_k}         = 2\nu \frac{\partial^2 \langle \delta u_i \delta u_j \rangle}{(\partial r_k)^2}.
        \label{eq:isotropicineqj}
\end{eqnarray}
When $i=j$, Equation \eqref{eq:isotropicij} reduces to \eqref{eq:isotropicsbs}. So the energy of different scales may transfer energy within the terms in Equation \eqref{eq:isotropicineqj}, but the energy only dissipates in the terms of Equation \eqref{eq:isotropicsbs}.

Equation \eqref{eq:isotropicsbs} is similar to Equation (47) in \cite{Karman1938}, but expressed in terms of velocity differences. When the mean velocity is homogeneous and $\delta U_i = 0$, Equation \eqref{eq:isotropicsbs} reduces to its homogeneous form, Equation \eqref{eq:homogeneousisotropicsbs}.

\section{The second form of the two-point equations}
Adding the two equations eliminates certain terms, such as pressure. It may seem counterintuitive that no pressure term influences the flow. However, if we subtract the equations instead, the remaining terms reveal the hidden influence that was canceled out in the addition process. So, the second form of the two-point equation can be obtained by subtracting the two equations in Equation \eqref{eq:RDdelta}:
\begin{eqnarray*}
        \displaystyle \left(\frac{\partial }{\partial t^+} - \frac{\partial }{\partial t^-}\right)(\delta U_i + \delta u_i)+ \frac{\partial (U_k^+ + u_k^+)(\delta U_i + \delta u_i) }{\partial x^+_k} -  \frac{\partial (U_k^- + u_k^-) (\delta U_i + \delta u_i)}{\partial x^-_k}
        \displaystyle \\[2ex]
        = - \left(\frac{\partial }{\partial x^+_i} -\frac{\partial }{\partial x^-_i} \right)(\delta P +\delta p) + \nu \left[\frac{\partial^2 }{(\partial x^+_k)^2}-\frac{\partial^2 }{(\partial x^-_k)^2}\right](\delta U_i + \delta u_i)
\end{eqnarray*}

Substituting the new variables \eqref{eq:newvaribles} into the equation above and simplifying yields
\begin{eqnarray}
        \displaystyle \frac{\partial }{\partial r_0} (\delta U_i + \delta u_i)+ 2\frac{\partial (U_k^* + u_k^*)(\delta U_i + \delta u_i)}{\partial r_k} +\frac{1}{2}\frac{\partial (\delta U_k + \delta u_k)(\delta U_i + \delta u_i)}{\partial X_k} \nonumber \\[2ex]
       \displaystyle  = - \frac{\partial }{\partial r_i} \left(\delta P +\delta p\right) ++ \nu \left[\frac{\partial^2 }{(\partial x^+_k)^2}-\frac{\partial^2 }{(\partial x^-_k)^2}\right](\delta U_i + \delta u_i)
       \label{eq:substract3}
\end{eqnarray}

Multiplying Equation \eqref{eq:substract3} by $\delta u_j$:
\begin{eqnarray}
        \displaystyle \delta u_j\frac{\partial }{\partial r_0} (\delta U_i + \delta u_i)+ 2\delta u_j\frac{\partial (U_k^* + u_k^*)(\delta U_i + \delta u_i)}{\partial r_k} +\frac{\delta u_j}{2}\frac{\partial (\delta U_k + \delta u_k)(\delta U_i + \delta u_i)}{\partial X_k} \nonumber \\[2ex]
       \displaystyle  = - \delta u_j \frac{\partial }{\partial r_i} (\delta P +\delta p) +        \nu \delta u_j \left[\frac{\partial^2 }{(\partial x^+_k)^2}-\frac{\partial^2 }{(\partial x^-_k)^2}\right](\delta U_i + \delta u_i) \quad \quad
       \label{eq:substractmultiplyij}
\end{eqnarray}

Setting $i=j$ in Equation \eqref{eq:substractmultiplyij} and multiplying the identity by 2, the summing over index $i$ results in the energy equation:
\begin{eqnarray}
        \displaystyle 2\delta u_i\frac{\partial \delta U_i}{\partial r_0} + \frac{\partial (\delta u_i)^2}{\partial r_0} + 4\delta u_i U_k^*\frac{\partial \delta U_i }{\partial r_k} + 4\delta u_i u_k^*\frac{\partial \delta U_i }{\partial r_k} + ( U_k^*+u_k^* )\frac{\partial (\delta u_i)^2}{\partial r_k} \nonumber \\[2ex]
        +\delta u_i \delta U_k\frac{\partial \delta U_i }{\partial X_k} + \delta u_i \delta u_k\frac{\partial\delta U_i }{\partial X_k} + \delta U_k\frac{\partial (\delta u_i)^2}{\partial X_k} + \delta u_k\frac{\partial (\delta u_i)^2}{\partial X_k}  \nonumber \\[2ex]
       \displaystyle  = - \frac{\partial (\delta u_i \delta P +\delta u_i \delta p)}{\partial r_i}  + \nu \delta u_i \left[\frac{\partial^2 }{(\partial x^+_k)^2}-\frac{\partial^2 }{(\partial x^-_k)^2}\right](\delta U_i + \delta u_i)
       \label{eq:substractmultiply}
\end{eqnarray}

Utilizing \eqref{appeq:viscosityij} in Appendix \ref{appB}, the viscous term in Equation \eqref{eq:substractmultiply} yields
\begin{eqnarray*}
    &\displaystyle 2\nu \delta u_j \left[\frac{\partial^2 }{(\partial x^+_k)^2}-\frac{\partial^2 }{(\partial x^-_k)^2}\right]\delta u_i \\[2ex]
    =&\displaystyle \nu \frac{\partial^2 \delta u_i \delta u_j}{\partial r_i\partial X_i} - 2\nu\frac{\partial \delta u^+_i}{\partial x^+_k}\frac{\partial \delta u^+_j}{\partial x^+_k}+ 2\nu\frac{\partial \delta u^-_i}{\partial x^-_k}\frac{\partial \delta u^-_j}{\partial x^-_k} 
\end{eqnarray*}

For $i=j$, the viscous term becomes
\begin{eqnarray}
    &\displaystyle 2\nu \delta u_i \left[\frac{\partial^2 }{(\partial x^+_k)^2}-\frac{\partial^2 }{(\partial x^-_k)^2}\right]\delta u_i \nonumber\\[2ex]
     =&\displaystyle \nu \frac{\partial^2 (\delta u_i)^2}{\partial r_i\partial X_i} - 2(\epsilon^+ - \epsilon^-) - 2\nu\Delta (\delta P -+  H^-)+ 2\nu\delta p + 2\nu F^+_{ij} -2\nu F^-_{ij},
     \label{eq:viscosityijsub}
\end{eqnarray}
where $\displaystyle H^-=\frac{\partial U^+_i}{\partial x^+_j}\frac{\partial U^+_j}{\partial x^+_i} - \frac{\partial U^-_i}{\partial x^-_j}\frac{\partial U^-_j}{\partial x^-_i}$. 

\subsection{The unaveraged form of the subtracted equation}
Equation \eqref{eq:substractmultiplyij} can be reproduced with the viscous terms \eqref{eq:viscosityijsub},
\begin{eqnarray}
        \displaystyle \delta u_j\frac{\partial \delta U_i}{\partial r_0} + \delta u_j\frac{\partial \delta u_i}{\partial r_0} + 2\delta u_j U_k^*\frac{\partial \delta U_i }{\partial r_k} + 2\delta u_j u_k^*\frac{\partial \delta U_i }{\partial r_k} + 2\delta u_j ( U_k^*+u_k^* )\frac{\partial \delta u_i}{\partial r_k} \nonumber \\[2ex]
        +\delta u_j \delta U_k\frac{\partial \delta U_i }{\partial X_k} + \delta u_j \delta u_k\frac{\partial\delta U_i }{\partial X_k} + \delta u_j\delta U_k\frac{\partial \delta u_i}{\partial X_k} + \delta u_j \delta u_k\frac{\partial \delta u_i}{\partial X_k}  \nonumber \\[2ex]
       \displaystyle  = - \frac{\partial (\delta u_i \delta P +\delta u_i \delta p)}{\partial r_i}  + \nu \frac{\partial^2 (\delta u_i)^2}{\partial r_i\partial X_i} - 2(\epsilon^+ - \epsilon^-)  \nonumber \\[2ex]
       - 2\nu\Delta (\delta P -+  H^-)+ 2\nu\delta p + 2\nu F^+_{ij} -2\nu F^-_{ij}
       \label{eq:substractmultiplyijunaverage}
\end{eqnarray}

For $i=j$, the second form of the unaveraged scale-by-scale equation is obtained, 
\begin{eqnarray}
        \displaystyle \underbrace{2\delta u_i\frac{\partial \delta U_i}{\partial r_0}}_{\substack{\text{Stochastic Tran-} \\ \text{sient transport}}} + \underbrace{\frac{\partial (\delta u_i)^2}{\partial r_0}}_{\substack{\text{Transient} \\ \text{TKE transfer}}} + \underbrace{2\delta u_i U_k^*\frac{\partial \delta U_i }{\partial r_k}}_{\substack{\text{Stochastic} \\ \text{production}}} + \underbrace{2\delta u_i u_k^*\frac{\partial \delta U_i }{\partial r_k}}_{\substack{\text{Turbulent prod-} \\ \text{uction by transfer}}} + \underbrace{ U_k^*\frac{\partial (\delta u_i)^2}{\partial r_k}}_{\substack{\text{Linear TKE} \\ \text{transfer}}}  \\[2ex]
        + \underbrace{u_k^* \frac{\partial (\delta u_i)^2}{\partial r_k}}_{\substack{\text{Nonlinear TKE} \\ \text{ transfer}}} + \underbrace{\delta u_i \delta U_k\frac{\partial \delta U_i }{\partial X_k}}_{\substack{\text{Stochastic} \\ \text{ transport}}} +\underbrace{\delta u_i \delta u_k\frac{\partial\delta U_i }{\partial X_k}}_{\substack{\text{Turbulent production} \\ \text{by transport}}} +\underbrace{\frac{ \delta U_k }{2}\frac{\partial (\delta u_i)^2}{\partial X_k}}_{\substack{\text{Linear TKE} \\ \text{transport}}} +\underbrace{\frac{ \delta u_k}{2}\frac{\partial (\delta u_i)^2}{\partial X_k}}_{\substack{\text{Nonlinear TKE} \\ \text{ transfer}}} \nonumber \\[2ex]
       \displaystyle  = - \underbrace{2\frac{\partial \delta u_i \delta P }{\partial r_i}}_{\substack{\text{Stochastic pres-} \\ \text{sure transfer}}} - \underbrace{2\frac{\partial \delta u_i \delta p}{\partial r_i}}_{\substack{\text{Turbulent Pres-} \\ \text{sure transfer}}} + \underbrace{2\nu \delta u_i \frac{\partial^2 \delta U_i}{\partial r_i\partial X_i}}_{\substack{\text{Stochastic viscous} \\ \text{transfer}}} +  \underbrace{2\nu \frac{\partial^2 (\delta u_i)^2}{\partial r_i\partial X_i}}_{\substack{\text{turbulent viscous} \\ \text{diffusion}}}\nonumber \\[2ex]
         \displaystyle- \underbrace{2(\epsilon^+ - \epsilon^-)}_{\substack{\text{Dissipation} \\ \text{difference}}} - \underbrace{2 \nu (\Delta \delta P -  2H^-)}_{\substack{\text{Viscous diffusion} \\ \text{transport from mean}}} - \underbrace{2\nu \Delta \delta p+ 2\nu F^+_{ij} -2\nu F^-_{ij}}_{\substack{\text{Stochastic viscous} \\ \text{diffusion}}}
         \label{eq:substractmultiplyinter}
\end{eqnarray}
All stochastic terms vanish in the averaging operation.

\subsection{The averaged form of the substracted scale by scale equation}
Phase averaging is applied to Equation \eqref{eq:substractmultiplyinter} to obtain the averaged form
\begin{eqnarray}
        \displaystyle \underbrace{\frac{\partial \langle (\delta u_i)^2\rangle}{\partial r_0}}_{\substack{\mathcal{A}'\text{ Transient} \\ \text{TKE transfer}}} + \underbrace{2\langle\delta u_i u_k^*\rangle\frac{\partial \delta U_i }{\partial r_k}}_{\substack{\mathcal{F}_1'\text{ Turbulent prod-} \\ \text{uction by transfer}}} + \underbrace{ U_k^*\frac{\partial \langle(\delta u_i)^2\rangle}{\partial r_k}}_{\substack{\mathcal{L}_1'\text{ Linear TKE} \\ \text{transfer}}}  \\[2ex]
        + \underbrace{\frac{\partial \langle u_k^* (\delta u_i)^2\rangle}{\partial r_k}}_{\substack{\mathcal{\Pi}_1'\text{ Nonlinear} \\ \text{TKE transfer}}} +\underbrace{\langle \delta u_i \delta u_k\rangle\frac{\partial\delta U_i }{\partial X_k}}_{\substack{\mathcal{F}_2'\text{ Turbulent produ-} \\ \text{ction by transport}}} +\underbrace{\frac{ \delta U_k }{2}\frac{\partial \langle (\delta u_i)^2\rangle}{\partial X_k}}_{\substack{\mathcal{L}_2'\text{ Linear TKE} \\ \text{transport}}} +\underbrace{\frac{1}{2}\frac{\partial \langle \delta u_k (\delta u_i)^2\rangle}{\partial X_k}}_{\substack{\mathcal{\Pi}_2'\text{ Nonlinear} \\ \text{TKE transport}}} \nonumber \\[2ex]
       \displaystyle  = - \underbrace{2\frac{\partial \langle\delta u_i \delta p\rangle}{\partial r_i}}_{\substack{\mathcal{P}'\text{ Turbulent Pres-} \\ \text{sure transfer}}} +  \underbrace{2\nu \frac{\partial^2 \langle(\delta u_i)^2\rangle}{\partial r_i\partial X_i}}_{\substack{\mathcal{D}_1'\text{ Turbulent viscous} \\ \text{diffusion}}} - \underbrace{2(\langle\epsilon^+\rangle - \langle\epsilon^-\rangle)}_{\substack{\mathcal{D}_\epsilon'\text{ Dissipation} \\ \text{difference}}} - \underbrace{2 \nu (\Delta \delta P -  2H^-)}_{\substack{\mathcal{D}_2'\text{Viscous diffusion} \\ \text{transport from mean}}}
         \label{eq:substractmultiplayaverage}
\end{eqnarray}

\subsection{The assumption of homogeneity}
The homogeneity assumption is employed to \eqref{eq:substractmultiplayaverage} resulting in
\begin{eqnarray}
        \displaystyle \frac{\partial \langle (\delta u_i)^2\rangle}{\partial r_0} + U_k^*\frac{\partial \langle(\delta u_i)^2\rangle}{\partial r_k} + \frac{\partial \langle u_k^* (\delta u_i)^2\rangle}{\partial r_k} +\langle \delta u_i \delta u_k\rangle\frac{\partial\delta U_i }{\partial X_k} = - 2\frac{\partial \langle\delta u_i \delta p\rangle}{\partial r_i}
        \label{eq:homogeneoussubstract}
\end{eqnarray}
Note that there is no dissipation term left in this equation, and the remaining terms (in symbol form) are
\begin{equation*}
    \mathcal{A}' + \mathcal{L}'_1 + \mathcal{\Pi}'_1 + \mathcal{F}'_2 = - \mathcal{P}'.
\end{equation*}

\subsection{The assumption of isotropy}
The isotropy assumption is applied to \eqref{eq:homogeneoussubstract} resulting in
\begin{eqnarray}
        \displaystyle \frac{\partial \langle (\delta u_i)^2\rangle}{\partial r_0} + U_k^*\frac{\partial \langle(\delta u_i)^2\rangle}{\partial r_k} + \frac{\partial \langle u_k^* (\delta u_i)^2\rangle}{\partial r_k} = - 2\frac{\partial \langle\delta u_i \delta p\rangle}{\partial r_i}
        \label{eq:isotropicsubstract}
\end{eqnarray}
The remaining terms (in symbol form) are:
\begin{equation*}
    \mathcal{A}' + \mathcal{L}'_1 + \mathcal{\Pi}'_1 = - \mathcal{P}'.
\end{equation*}
This form of the equation can be used for analyzing energetic nonlinear interactions with transient behaviour and pressure effects, both of which do not exist in the scale-by-scale equation \eqref{eq:homogeneousisotropicsbs}. From Equation \eqref{eq:homogeneoussubstract} and \eqref{eq:isotropicsubstract}, we can conclude that the pressure term does not contribute directly to dissipation. However, it plays a role in the scale-by-scale energy transfer, affecting both spatial and temporal dynamics.
\section{Conclusion and discussion}
We have conducted an overview of the development of the KHMH equations in the literature, revealing discrepancies in their application. We provided the exact two-point (spatio-temporal) structure equation using Reynolds decomposition, along with its phase average form.

Under assumptions of homogeneity and isotropy, we presented a simplified equation closely resembling those proposed in seminal works such as \cite{Karman1938}, \cite{Monin1971} and \cite{Kolmogorov1941c}. The equations can be used in the scale-by-scale energy cascade analysis and nonlinear energy transfer like \cite{Chen2022, Valente2015, Gomes2015,Alves2017, Alves2020, Yao2022, Yao2023,Valente2015}.

Furthermore, we derived a new form of the scale-by-scale equation incorporating pressure contributions and mean velocity production. This formulation paves the way for discussions with researchers interested in exploring interscale interactions involving pressure terms.

\appendix
\section{The Discrepancies of scale-by-scale equations}\label{appA}
In their seminal work, \cite{Karman1938} initiated the development of the scale-by-scale equation. They began with Equation (41) to Equation (50) (in \cite{Karman1938}), which employed the two-point Navier-Stokes equations. Multiplying the velocity of one point to the equation of the opposite point and subsequently averaging it, and following af process involving summing the two equations while incorporating several average properties of isotropic turbulence led to the following relation:
\begin{equation}
    \overline{ \frac{\partial u_i u_k'}{\partial t}}-\frac{\partial}{\partial \xi_j}(\overline{u_i u_j u_k'}+\overline{u_i' u_j u_k}) = 2\nu \nabla^2(\overline{u_i u_k'}).
\end{equation}
With the distance function under the isotropic assumption, Equation (51) is obtained as
\begin{equation}\label{eqapp:karman}
    \frac{\partial f\overline{u^2}}{\partial t} + 2(\overline{u^2})^{\frac{3}{2}}\left( \frac{\partial h}{\partial r} + \frac{4}{r}h\right) = 2\nu \overline{u^2} \left( \frac{\partial^2 f}{\partial r^2} + \frac{4}{r}\frac{\partial f}{\partial r}\right),
\end{equation}
where $f$ is the distance function in the two-point correlation matrix, and $(\overline{u^2})^{\frac{3}{2}}h=\overline{u_2^2u_1'}=\overline{u_3^2u_1'} = -\overline{u_1^2u_1'}/2$.

Kolmogorov declared the utilization of this equation, albeit with a focus on the statistics of the velocity difference between the two points. This was articulated in Equation (3) of \cite{Kolmogorov1941c}.
\begin{equation}\label{eqapp:kolmo}
    4\overline{E} + \left( \frac{\partial B_{ddd}}{\partial r} + \frac{4}{r}B_{ddd}\right) = 6\nu \left( \frac{\partial^2 B_{dd}}{\partial r^2} + \frac{4}{r}\frac{\partial B_{dd}}{\partial r}\right),
\end{equation}
where $B_{ddd}=\overline{[u_d(M')-u_d(M))]^3}$ is the third-order moment of the velocity components in the direction $\overline{MM'}$, and $B_{dd}=\overline{[u_d(M')-u_d(M))]^2}$ is the second-order moment of the velocity components in the direction $\overline{MM'}$, and $\overline{E}$ denotes the mean dissipation rate of energy. Then Kolmogorov yields
\begin{equation}\label{eqapp:kolmosolution}
    6\nu \frac{\partial B_{dd}}{\partial r} - B_{ddd}  = \frac{4}{5}\overline{E}r
\end{equation}

The disparity between Equation \eqref{eqapp:karman} and \eqref{eqapp:kolmo} is evident. This relationship can be expressed as follows: $B_{dd}= \overline{u^2}f/3$, $B_{ddd}=2(\overline{u^2})^{\frac{3}{2}}h$, and $\overline{E}= \frac{1}{4} \frac{\partial f\overline{u^2}}{\partial t}$, where $\overline{E}$ denotes the mean dissipation rate of energy per unit of time per unit of mass.

While Kolmogorov did not explicitly detail the transformation of the equation from the velocity form to the velocity difference form (or statistical moments form), Monin undertook a reformulation of the equation with the velocity difference of two points. This reformulation followed the framework established in \cite{Karman1938}, utilizing the difference in velocity in the context of anisotropic turbulence. This process is elaborated upon from page 401 to page 403, specifically in equation 22.16 (\cite{Monin1971}):
\begin{equation}\label{eqapp:Monin}
     \frac{\partial  D_{ijk}}{\partial r_k}  =  2\nu\Delta D_{ij} - \frac{4}{3} \overline{\epsilon} \delta_{ij},
\end{equation}
where the second-order moments $D_{ij}=\overline{(u_i(x+r)-u_i(x))(u_j(x+r)-u_j(x))}$ and the third-order moments $D_{ijk}=\overline{(u_i(x+r)-u_i(x))(u_j(x+r)-u_j(x))(u_k(x+r)-u_k(x))}$, and following \cite{Kolmogorov1941c} obtained as, the solution is
\begin{equation}\label{eqapp:Moninsolution}
     D_{LLL}(r) -6\nu  \frac{\partial D_{LL}(r)}{\partial r} = - \frac{4}{5} \overline{\epsilon}r+ \frac{C_1}{r^2}+\frac{C_2}{ r^4}.
\end{equation}
With local isotropy assumption and , Equation \eqref{eqapp:Moninsolution} can be reduced to Kolmogorov's conclusion in Equation \eqref{eqapp:kolmosolution} (see section 22.1 and section 13.3 in \cite{Monin1971}).

Frisch attributed the development of the anisotropic version of the equation "largely" to Monin, particularly referencing Moin's work in 1959 \cite{Frisch1995}. However, Danaila referred to it as the 'Yaglom-like equation' and reformulated it based on the two variables' equation outlined in \cite{Hill2001, Hill2002}. The results obtained from Equation (14) in \cite{Danaila2012} appear ambiguous due to their utilization of both old and new variables. It is worth noting that Equation (9) in \cite{Danaila2012} does not align with the second line of Equation (1) in \cite{Valente2015}, as both pertain to the second point in the two-point equation, raising skepticism about the correctness of Equation (10) in \cite{Danaila2012}.

From the beginning of the scale-by-scale equation, the same instance in time is being considered for the two points in the equation. \cite{Monin1971} suggested the spatio-temporal two-point in the statistics studies in turbulence (see Section 21.1 in \cite{Monin1971}). In the works of \cite{Hill2001, Hill2002}, the spatial two-point equation was deduced without utilizing Reynolds decomposition. \cite{Yasuda2018} employed this equation to analyze turbulence intermittency in DNS data.

On the other hand, \cite{Valente2015} derived the spatial two-point equation using the Reynolds decomposition and time averaging. The last viscous term in Equation (2) was interpreted as the dissipation rate without any assumptions. However, it is noted that the term $2\nu\langle \left( \partial u_i/\partial x_j\right)^2\rangle$ represents the dissipation rate only for homogeneous flows (see Section 9.7 in \cite{George2013}). A similar dissipation interpretation can be found in \cite{Valente2015, Gomes2015, Alves2017, Alves2020, Yao2022}.

Throughout the evolution of the scale-by-scale equation, a consistent unique instance in time step has been utilized for both spatially separated points across these equations.
\section{Mathematical properties}\label{appB}
The two points are independent. The derivatives applied of the two points thus have the following properties:
\begin{equation}\label{eq:independent1}
    \begin{cases}
        \displaystyle \frac{\partial U_i^- }{\partial x^+_j} = \frac{\partial u_i^- }{\partial x^+_j}=  \frac{\partial P^- }{\partial x^+_j} = \frac{\partial p^- }{\partial x^+_j} = \nu \frac{\partial^2 U_i^- }{\partial x^+_j}= \nu \frac{\partial^2 u_i^-}{\partial x^+_j} = 0\\[2ex] 
        \displaystyle\frac{\partial U_i^+ }{\partial x^-_j} = \frac{\partial u_i^+ }{\partial x^-_j} = \frac{\partial P^+ }{\partial x^-_j} = \frac{\partial p^+ }{\partial x^-_j} = \nu \frac{\partial^2 U_i^+ }{\partial x^-_j}= \nu \frac{\partial^2 u_i^+}{\partial x^-_j} = 0
    \end{cases}.
\end{equation}
Here, the indices $i,j$ do not mean summation in tensor notation. Furthermore, $(U_k^+ + u_k^+)\frac{\partial U_i^- }{\partial x^+_j} = (U_k^+ + u_k^+)\frac{\partial u_i^- }{\partial x^+_j} =0$.

So, the partial derivatives of the two new variables in both the first order and the second order are obtainable by defining a linear matrix:
\begin{equation}\label{appeq:M1}
    \begin{bmatrix}
    \displaystyle\frac{\partial}{\partial r_i}\\[2ex]
   \displaystyle \frac{\partial}{\partial X_i}\\[2ex]
    \displaystyle\frac{\partial^2}{\partial r^2_i}\\[2ex]
    \displaystyle\frac{\partial^2}{\partial r_i \partial X_i}\\[2ex]
    \displaystyle\frac{\partial^2}{\partial X^2_i}\\[2ex]
    \end{bmatrix}
    =\mathbf{M}
    \begin{bmatrix}
    \displaystyle\frac{\partial}{\partial x^+_i}\\[2ex]
    \displaystyle\frac{\partial}{\partial x^-_i}\\[2ex]
    \displaystyle\frac{\partial^2}{(\partial x^+_i)^2}\\[2ex]
    \displaystyle\frac{\partial^2}{\partial x^+_i \partial x^-_i}\\[2ex]
    \displaystyle\frac{\partial^2}{(\partial x^-_i)^2}\\[2ex]
    \end{bmatrix}.
\end{equation}

The matrix $\mathbf{M}$ can be obtained by multiplying row vector $[r_i, X_i, r_i^2, r_iX_i, X^2_i]$ from the left side in both sides of the equation; then the matrix can be solved yielding:

\begin{equation}\label{appeq:M2}
    \mathbf{M}=
    \begin{bmatrix}
    \displaystyle \frac{1}{2} & \displaystyle-\frac{1}{2} & 0 & 0 & 0\\[2ex]
     1 & 1 & 0 & 0 & 0\\[2ex]
     0 & 0 & \displaystyle\frac{1}{4} & \displaystyle-\frac{1}{2} & \displaystyle\frac{1}{4}\\[2ex]
     0 & 0 & \displaystyle\frac{1}{2} & 0 & \displaystyle-\frac{1}{2}\\[2ex]
     0 & 0 & 1 & 2 & 1\\[2ex]
    \end{bmatrix}
    ,\quad \mathbf{M}^{-1}=
    \begin{bmatrix}
     1 & \displaystyle\frac{1}{2} & 0 & 0 & 0\\[2ex]
     -1 & \displaystyle\frac{1}{2} & 0 & 0 & 0\\[2ex]
     0 & 0 & 1 & 1 & \displaystyle\frac{1}{4}\\[2ex]
     0 & 0 & -1 & 0 & \displaystyle\frac{1}{4}\\[2ex]
     0 & 0 & 1 & -1 & \displaystyle\frac{1}{4}\\[2ex]
    \end{bmatrix}
\end{equation}
Then equations \eqref{eq:newvaribles} and \eqref{eq:2rdnewvaribles} can be obtained.

From the continuity equation of incompressible flow, $\displaystyle \frac{\partial U^+_k}{\partial x^+_k} = \frac{\partial u^+_k}{\partial x^+_k}= \frac{\partial U^-_k}{\partial x^-_k} = \frac{\partial u^-_k}{\partial x^-_k} = 0$, the nonlinear term in the two-point equation can be expressed as:
\begin{eqnarray}\label{appeq:nonlinear1}
        \displaystyle (U_k^+ + u_k^+)\frac{\partial }{\partial x^+_k}(\delta U_i + \delta u_i) - (U_k^- + u_k^-) \frac{\partial}{\partial x^-_k}  (\delta U_i + \delta u_i)
        \displaystyle \nonumber \\[2ex] 
        = \displaystyle \frac{\partial (U_k^+ + u_k^+)(\delta U_i + \delta u_i) }{\partial x^+_k} -  \frac{\partial (U_k^- + u_k^-) (\delta U_i + \delta u_i)}{\partial x^-_k} 
\end{eqnarray}

Furthermore, the new variables also have the properties
\begin{equation}
\begin{cases}
    \displaystyle\frac{\partial U^+_k}{\partial r_k} = \frac{\partial u^+_k}{\partial r_k}= \frac{\partial U^-_k}{\partial r_k} = \frac{\partial u^-_k}{\partial r_k} = 0\\[2ex]
    \displaystyle\frac{\partial U^+_k}{\partial X_k} = \frac{\partial u^+_k}{\partial X_k}= \frac{\partial U^-_k}{\partial X_k} = \frac{\partial u^-_k}{\partial X_k} = 0
    \end{cases}.
\end{equation}
So the new velocity variables also have the following properties:

\begin{equation}
    \begin{cases}
    \displaystyle\frac{\partial U^*_k}{\partial r_k} = \frac{\partial u^*_k}{\partial r_k}= \frac{\partial \delta U_k}{\partial r_k} = \frac{\partial \delta u_k}{\partial r_k} = 0\\[2ex]
    \displaystyle\frac{\partial U^*_k}{\partial X_k} = \frac{\partial u^*_k}{\partial X_k}= \frac{\partial \delta U_k}{\partial X_k} = \frac{\partial \delta u_k}{\partial X_k} = 0
    \end{cases}
\end{equation}
Here, the index $k$ implies tensor notation and summation.

The second-order derivatives have the known property
\begin{equation*}
    \frac{\partial^2 f\cdot g}{\partial x^2_i} = f\frac{\partial^2 g}{\partial x^2_i} + g\frac{\partial^2 f}{\partial x^2_i} + 2\frac{\partial f}{\partial x_i}\frac{\partial g}{\partial x_i}
\end{equation*}
Similarly 
\begin{equation}
    \frac{\partial^2 f\cdot f}{\partial x^2_i} =2 f\frac{\partial^2 f}{\partial x^2_i} + 2\frac{\partial f}{\partial x_i}\frac{\partial f}{\partial x_i}
\end{equation}
The second-order derivative can be used in the velocity form,
\begin{equation*}
\begin{cases}
     \displaystyle\delta u_j\frac{\partial^2 \delta u_i}{(\partial x^+_k)^2}  = \displaystyle\frac{\partial^2 \delta u_i \delta u_j}{(\partial x^+_k)^2} - \delta u_i\frac{\partial^2 \delta u_j}{(\partial x^+_k)^2} - 2 \frac{\partial \delta u_i}{\partial x^+_k}\frac{\partial \delta u_j}{\partial x^+_k}\\[2ex]
     \displaystyle\delta u_j\frac{\partial^2 \delta u_i}{(\partial x^-_k)^2}  = \displaystyle\frac{\partial^2 \delta u_i \delta u_j}{(\partial x^-_k)^2} - \delta u_i\frac{\partial^2 \delta u_j}{(\partial x^-_k)^2} - 2 \frac{\partial \delta u_i}{\partial x^-_k}\frac{\partial \delta u_j}{\partial x^-_k}
\end{cases}
\end{equation*}
Here the index $i,j$ can commute in the summation, and the expression can be formulated as
\begin{equation}\label{appeq:viscosityij}
\begin{cases}
     \displaystyle\delta u_j\frac{\partial^2 \delta u_i}{(\partial x^+_k)^2} = \displaystyle \frac{1}{2}\frac{\partial^2 \delta u_i \delta u_j}{(\partial x^+_k)^2} - \frac{\partial \delta u_i}{\partial x^+_k}\frac{\partial \delta u_j}{\partial x^+_k}\\[2ex]
     \displaystyle \delta u_j\frac{\partial^2 \delta u_i}{(\partial x^-_k)^2} = \displaystyle \frac{1}{2}\frac{\partial^2 \delta u_i \delta u_j}{(\partial x^-_k)^2} - \frac{\partial \delta u_i}{\partial x^-_k}\frac{\partial \delta u_j}{\partial x^-_k}
\end{cases}
\end{equation}

When $i=j$, this term reduces to
\begin{equation}\label{appeq:viscosityii}
\begin{cases}
     \displaystyle \delta u_i\frac{\partial^2 \delta u_i}{(\partial x^+_k)^2} = \displaystyle\frac{1}{2}\frac{\partial^2 \delta u_i \delta u_i}{(\partial x^+_k)^2} - \frac{\partial \delta u_i}{\partial x^+_k}\frac{\partial \delta u_i}{\partial x^+_k}.\\[2ex]
     \displaystyle \delta u_i\frac{\partial^2 \delta u_i}{(\partial x^-_k)^2} = \displaystyle\frac{1}{2}\frac{\partial^2 \delta u_i \delta u_i}{(\partial x^-_k)^2} - \frac{\partial \delta u_i}{\partial x^-_k}\frac{\partial \delta u_i}{\partial x^-_k}.
\end{cases}
\end{equation}

The Poisson equation yields
\begin{eqnarray*}
    -\Delta (P+p)& = & \displaystyle\frac{\partial (U_i+u_i)}{\partial x_j }\frac{\partial (U_j+u_j)}{\partial x_i} \nonumber\\[2ex]
    & = & \displaystyle\frac{\partial U_i}{\partial x_j}\frac{\partial U_j}{\partial x_i}+\frac{\partial u_i}{\partial x_j}\frac{\partial U_j}{\partial x_i}+\frac{\partial U_i}{\partial x_j}\frac{\partial u_j}{\partial x_i}+\frac{\partial u_i}{\partial x_j}\frac{\partial u_j}{\partial x_i}
\end{eqnarray*}
We take the form $\displaystyle -\Delta (P+p) = F_{ij} + \frac{\partial U_i}{\partial x_j}\frac{\partial U_j}{\partial x_i}+ \frac{\partial u_i}{\partial x_j}\frac{\partial u_j}{\partial x_i}$, as $F_{ij} = \displaystyle\frac{\partial u_i}{\partial x_j}\frac{\partial U_j}{\partial x_i}+\frac{\partial U_i}{\partial x_j}\frac{\partial u_j}{\partial x_i}$.
While the fluctuating part of the strain rate with Reynolds stress is $s_{ij} = \displaystyle \frac{1}{2} \left( \frac{\partial u_i}{\partial x_j}+\frac{\partial u_j}{\partial x_i}\right)$, and the instantaneous dissipation rate is $\epsilon = 2\nu s_{ij}s_{ij}$, the indces $i,j$ can be commuted and the residual term in equation \eqref{appeq:viscosityii} equals to
\begin{equation}
     \left(\frac{\partial u_i}{\partial x_j}\right)^2 = \nu^{-1}\epsilon + \Delta (P+p) + F_{ij} +\frac{\partial U_i}{\partial x_j}\frac{\partial U_j}{\partial x_i}
     \label{appeq:viscous}
\end{equation}
Expand the mean velocity transfer term $H^+ $ with $ab+cd = \frac{1}{2}[(a+c)(b+d)+(a-c)(b-d)]$
\begin{eqnarray}
    \frac{\partial U^+_i}{\partial x^+_j}\frac{\partial U^+_j}{\partial x^+_i}+\frac{\partial U^-_i}{\partial x^-_j}\frac{\partial U^-_j}{\partial x^-_i}=\frac{\partial \delta U_i}{\partial x^+_j}\frac{\partial \delta U_j}{\partial x^+_i}+\frac{\partial \delta U_i}{\partial x^-_j}\frac{\partial \delta U_j}{\partial x^-_i} \nonumber\\[2ex]
    = \frac{1}{2}\left[ (\frac{\partial \delta U_i}{\partial x^+_j} + \frac{\partial \delta U_i}{\partial x^-_j})(\frac{\partial \delta U_j}{\partial x^+_i} + \frac{\partial \delta U_j}{\partial x^-_i}) + (\frac{\partial \delta U_i}{\partial x^+_j} - \frac{\partial \delta U_i}{\partial x^-_j})(\frac{\partial \delta U_j}{\partial x^+_i} - \frac{\partial \delta U_j}{\partial x^-_i})\right]\nonumber\\[2ex]
    = \frac{1}{2}\frac{\partial \delta U_i}{\partial X_j} \frac{\partial \delta U_j}{\partial X_i} + 2\frac{\partial \delta U_i}{\partial r_j} \frac{\partial \delta U_j}{\partial r_i}.
\end{eqnarray}

Expand the mean velocity transfer term $H^- $ with $ab-cd = \frac{1}{2}[(a+c)(b-d)+(a-c)(b+d)]$
\begin{eqnarray}
    \frac{\partial U^+_i}{\partial x^+_j}\frac{\partial U^+_j}{\partial x^+_i} - \frac{\partial U^-_i}{\partial x^-_j}\frac{\partial U^-_j}{\partial x^-_i}=\frac{\partial \delta U_i}{\partial x^+_j}\frac{\partial \delta U_j}{\partial x^+_i} - \frac{\partial \delta U_i}{\partial x^-_j}\frac{\partial \delta U_j}{\partial x^-_i}\nonumber\\[2ex]
    = \frac{1}{2}\left[ (\frac{\partial \delta U_i}{\partial x^+_j} + \frac{\partial \delta U_i}{\partial x^-_j})(\frac{\partial \delta U_j}{\partial x^+_i}-\frac{\partial \delta U_j}{\partial x^-_i}) + (\frac{\partial \delta U_i}{\partial x^+_j} - \frac{\partial \delta U_i}{\partial x^-_j})(\frac{\partial \delta U_j}{\partial x^+_i}+\frac{\partial \delta U_j}{\partial x^-_i})\right]\nonumber\\[2ex]
    = \frac{\partial \delta U_i}{\partial X_j} \frac{\partial \delta U_j}{\partial r_i} + \frac{\partial \delta U_i}{\partial r_j} \frac{\partial \delta U_j}{\partial X_i}.
\end{eqnarray}


\bibliographystyle{plain}
\bibliography{KHMHbib}

\end{document}